# Site preference and diffusion behaviors of H influenced by the implanted-He in 3C-β SiC

*Sen Wang[†], H. Y. He[†*], R. Ding[§], B. C. Pan[†‡], and J. L. Chen[§]*

[†] Key Laboratory of Strongly-Coupled Quantum Matter Physics, Department of Physics,

[§] Institute of Plasma Physics, Chinese Academy of Sciences, Hefei 230031, People's Republic of China

[‡] Hefei National Laboratory for Physical Sciences at Microscale,

University of Science and Technology of China, Hefei, Anhui 230026, People's Republic of China

*Corresponding authors: e-mail hyhe@ustc.edu.cn, Phone: +86 0551 63600697, Fax: +86 0551 63601073

**Abstact:**

  SiC materials are potential plasma facing materials in fusion reactors. In this study, site preference and diffusion behaviors of H in pure 3C-β SiC and in He-implanted 3C-β SiC are investigated, on the basis of the first-principles calculations. We find that the most stable sites for H in pure 3C-βSiC is the anti-bond site of C (ABc) in Si-C, while it becomes the bond-center (BC) site of Si-C bonds in the He-implanted 3C-β SiC. Analysis on the electronic structures reveals that such change is attributed to the reduction of hybridization of C-Si bonds induced by He. Moreover, the presence of He strongly affect the vibrational features in the high frequency region, causing a blue shift of 25 cm$^{-1}$ for C-H stretch mode with H at ABc site and a red shift of 165cm$^{-1}$ for that at BC site, with respect to that in the pure system. In pure 3C-β SiC, H is diffusive with an energy cost of about 0.5 eV, preferring to rotate around the C atom in a Si-C tetrahedron with an energy barrier of just about 0.10 eV. In contrast, in He-implanted 3C-β SiC, the energy barriers for H migration goes up to be about 0.95 eV, indicating the implanted-He blocks the diffusive H to some extent. Our calculations also show that the influence of He on H diffusion is effective in a short range, just covering the nearest neighbor.

1. Introduction

SiC materials, possessing low induced-activation/low after-heat properties, creep resistance, especially exhibiting remarkable stability in high radiation environments, have been proposed as the prime candidates for plasma-facing materials in future fusion reactors [1-4]. Actually, SiC has recently been applied in the divertor of EAST [5-6]. Among various phases, the cubic beta-phase SiC (3C-β SiC) is the preferred material over the hexagonal alpha-phase SiC-based materials, because the latter in general experiences anisotropic swelling under irradiation, causing eventual intergranular cracking [4].

In a fusion process, both structural damage and property degradation were observed in 3C-β SiC [7-10]. For instance, the transmission electron microscopy showed that significant amounts of helium (He) retained in 3C-β SiC, and the growth of helium clusters induced the swelling of the system to some extent [11]. O'Connell et al. reported that a number of smaller bubbles and the loop-like defects were observed in the H-implanted 3C-β SiC. They proposed the loop-like defects are attributed to the mobile H diffusing from the main damaged region deeper into SiC during the annealing process [12]. Hydrogen isotope behaviors in 3C-β SiC were studied by Oya et al., and it was found that the deuterium (D) only bound to C in SiC[13]. Recently, the effect of He pre-irradiation on the H retention in 3C-β SiC was paid attention [14-15]. It was revealed that He irradiation lead to heavy damage in the host materials as compared to H irradiation, and the distribution of the retained H was observed not to be affected by preceding He implantation in 3C-β SiC[14].

On the theoretical side, the retention features of H and He in 3C-β SiC were focused on [16-17]. Gail and Son reported that the most stable site for H in pure 3C-β SiC is the anti-bond site of the C-Si bond, like H-C-Si [18-20]. Eddin et al. studied the insertion and diffusion of He in 3C-β SiC, and found that the most stable site of He in SiC is the tetrahedral interstitial site of four neighbouring Si (Tsi). They predicted the activation energies for migration in and around vacancies (silicon vacancy, carbon vacancy or divacancy) ranging from 0.6 to 1.0 eV [21-23]. It was found that diffusion barriers of He in pristine 3C-β SiC are about 1.05 eV and 1.55 eV for the paths of Tc-Tsi and Tsi-Tc, respectively, while those in H-implanted 3C-β SiC decrease to be 0.26 and 0.68 eV [24], much lower than the related values in the pure system. However, few study on diffusion behaviors of H in 3C-β SiC and that in He-implanted 3C-β SiC have been reported to our best knowledge.

In this work, we explore the site preference and diffusion behaviors of H in the pure 3C-β SiC, as well as in the He-implanted 3C-β SiC, on the basis of the density functional theory (DFT) calculations. We find that H favorably locates at the anti-bond site of Si-C (from the side of C) in the pure 3C-β SiC, whereas it prefers to stay at the bond center site of Si-C in the He-implanted 3C-β SiC. Moreover, the implanted He blocks the migration of H to some extent. The local vibrational features of H influenced by the implanted-He are addressed.

## 2. Computational method

All calculations are performed by using Vienna ab initio simulation package (VASP 5.3.3) based on the local density approximation [25]. The interaction between ions and valence electrons is described with using projector augmented wave (PAW) potentials[26], and the Perdew-Burke-Ernzerh of GGA expression is employed for the exchange-correlation functional[27]. A 64-atom supercell consisting of $2\times2\times2$ primitive unit cells is employed, and the kinetic energy cutoff of 500 eV is used for all calculations. In addition, the k–point sampling of $4\times4\times4$ within the Monkhorst-Pack special k-point scheme in the Brillouin zone is considered [28]. The energy relaxation iterates until the forces acting on all the atoms are less than $10^{-3}$ eV/Å. With these settings, we obtain the optimal crystallographic parameters of 3C-SiC, c=4.379 Å, which are in good agreement with the experimental values (4.360 Å) [29]. Since H is a light-mass particle, the zero-point energy (ZPE) of H is taken into account in the calculation [30]. In the present work, the ZPE of H atom is calculated by summing up the normal model of each H atom, $ZPE = \sum_i \eta \upsilon_i / 2$, where $\upsilon_i$ is the ith normal mode frequency.

The climbing image nudged elastic band (CI-NEB) method is used to determine the minimum energy paths for diffusion of H atoms [31-33] in SiC.

## 3. Results and discussion

### 3.1 Site preference of H in 3C-β SiC and in the He-implanted 3C-β SiC

We begin in exploring the favorite site of H in 3C-β SiC (for convenience, we just call it as SiC in the following). Four typical sites in SiC are considered, and they are the bond-center (BC) site of Si-C, the antibond site of Si-C (AB), the tetrahedral interstitial site of four neighboring Si (Tsi), and the tetrahedral interstitial site of four neighboring C(Tc), as shown in Fig.1. The formation energy of H at a site is defined as

$$E_f = E_s - E_{SiC}^{bulk} - E_H^{atom},$$

where $E_s$ is the total energy of the system containing H, $E_{SiC}^{bulk}$ is the total energy of the related pure SiC, and $E_H^{atom}$ is the energy of a single H atom. Of these sites, anti-bond of Si-C at the side of Si (ABsi) for H is not stable, at which H will move to Tsi spontaneously. Table 1 lists the calculated formation energy of H at the considered sites, in which the ZPE has been taken into account. From Table 1, one can see that the anti-bond site of Si-C at the side of C (ABc) is the energetically favorite site for H in the pure SiC, and the bond-center site of Si-C is also a preference site, at which the formation energy is only 0.06 eV higher than that at ABc site. This is in good consistence with previous reports [34-35]. We then consider the influence of He retention on the site preference of H in SiC. By introducing He into the pure SiC at various sites, we find the favorite site for He in SiC is the Tsi interstitial site, being in consistent with the literatures [21-23]. For convenience, we denote the He-implanted SiC as SiC+He. Similarly, the four typical sites for H in the pure SiC mentioned above are also taken into account around the implanted He (see Fig.1), and the calculated formation energies ($E_f'$) are summarized in Table 1. As seen in Table 1, the most stable site for H is not ABc site but the nearest BC site around He, with the formation energy being lower than that at ABc site by about 0.39eV. Moreover, H at the Tc site is not stable, due to the presence of He around it. The H atom at Tc site moves to ABc site spontaneously. Interestingly, our calculated formation energies of H at the next neighbouring concerned sites around He are almost the same as those in the pure SiC. This means that the presence of He in SiC can alter the distribution features of H within the scope of the neighboring region, which gets well with the experimental observation that the retained H distributions can be affected slightly by preceding He implantation in 3C-β SiC[14].

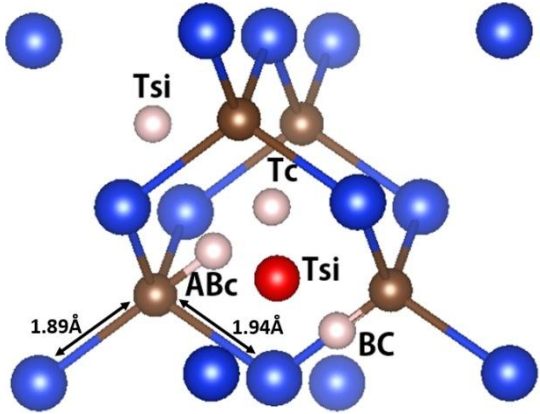

Fig.1 Local geometric structures for H at the sites of ABc, BC, Tsi and Tc in SiC. Blue, brown, red, pink balls represent Si, C, He, H atoms, respectively.

Table 1. Formation energy of H in SiC($E_f$) and SiC+He($E_f'$)

|            | ABc   | BC    | Tsi   | Tc    |
|---|---|---|---|---|
| $E_f$ (eV)  | 0.530 | 0.590 | 0.847 | 1.558 |
| $E_f'$ (eV) | 0.506 | 0.112 | 0.838 | --    |

We then discuss why the implanted He influences the site preference of H in SiC. By checking the configuration of SiC, we find that in the pure SiC system, the C-Si bond is about 1.89 Å in length, which is in good agreement with the experimental values (1.89 Å) [36]. With introduction of He into the system, the nearest C-Si bond around the He atom is enlarged to be 1.94 Å, while the next nearest neighboring C-Si bond also keeps its bond length of 1.89 Å, as shown in Fig.1. Careful examination of local structure around He reveals that the He atom at the site Tsi induces the displacement of the nearest Si atoms, and then the nearest neighboring C-Si bond becomes longer and consequently weaker. We further calculate the density of states (DOS) of the systems of both SiC and SiC+He, as displayed in Fig.2, in which the s-, p-projected DOS of the C、Si atoms in the pure SiC are shown in Fig.2 (a), and the s-, p-projected DOS of the C、Si atoms (the nearest neighbouring sites around the He atom) in the SiC+He are shown in Fig.2 (b). It clearly displays that the hybridization between p states of C and s states of Si atoms at the energy of Ef-7.2eV and Ef+4.8eV in Fig.2 (b) is smaller than that in Fig.2 (a), indicating weaker interaction between C and Si induced by He. As a result, H prefers to locate at the bond center site of C-Si in the He-implanted SiC, and consequently the formation energy of H in the nearest neighboring BC sites around the He atom decreases strikingly.

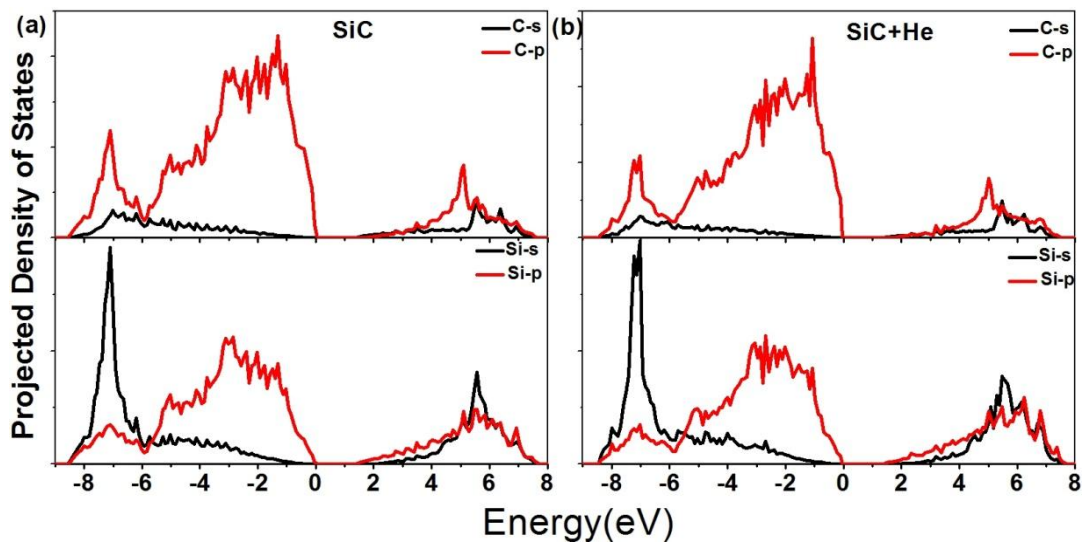

Fig.2 (a)The s-, p-projected Density of States (DOS) of the C、Si atoms in the pure SiC ; (b) The s-, p-projected DOS of the C、Si atoms(the nearest neighbour site around the He atom) in SiC+He.

Presence of He alters the site preference of H in SiC from site ABc to site BC, which should be reflected in vibration features. We now calculate the vibrational frequencies of these two systems (SiC and SiC+He) with H at both site ABc and site BC respectively by using the "frozen phonon" approach [37]. As shown in Fig.3 (a), compared with that in the pure SiC, the vibrational density of states (VDOS) of SiC with H at the site ABc have a significant new feature, in which a special peak appears at 2735 $cm^{-1}$. In contrast, for the case of SiC with H at the site BC, a new peak appears at 3075 $cm^{-1}$, with a blue shift of about 340 $cm^{-1}$ as compared to that in the former case. The calculated local vibrational density of states (LVDOS) [Fig.2 (b)] indicates that these two peaks correspond to the stretch vibrational modes of C-H in SiC, and the blue shift is attributed to the shortness of bond length（1.09 Å）of C-H at the site BC than that（1.12 Å）at the site ABc. Since the formation energy difference of H between at the site ABc and at the site BC in SiC is very small (less than 0.1 eV), we speculate these two kinds of stretch modes of C-H may coexist in vibrational spectrum of SiC. When He is introduced into SiC, we find that the special peaks corresponding to the C-H stretch mode shift to 2760 and 2910 $cm^{-1}$ for the cases of H at the sites of ABc and BC around He respectively. Compared with that in the pure SiC, the C-H stretch mode for SiC+He with H at the site ABc has a blue shift of about 25 $cm^{-1}$, and that at site BC has a red shift of 165 $cm^{-1}$. Such large red shift implies the C-H bond（1.11 Å）in the case with H at site BC is enlarged significantly, due to the presence of He.

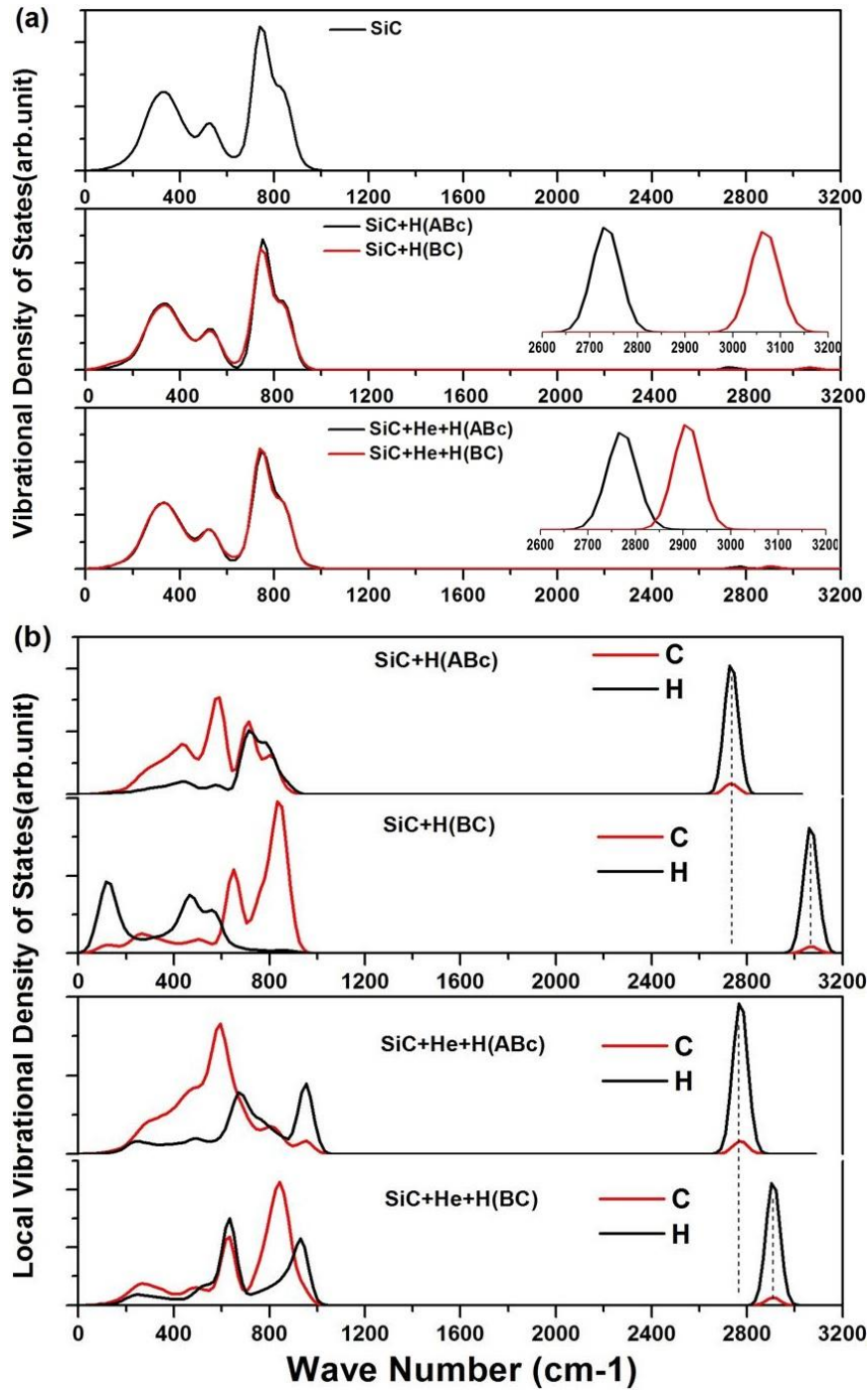

Fig.3 Vibrational density of states (a) of pure SiC, SiC+H with H at sites of ABc and BC, and SiC+He+H with H at sites of ABc and BC. (b) Local vibrational density of states of the nearest neighbouring H and C around the He atom.

## 3.2 H diffusion in SiC and SiC+He

After investigating the site preference, we turn to the diffusion behaviors of H in both SiC and SiC+He respectively. As we know, SiC is stacked up with Si-C

tetrahedrons(C and its four neighboring Si), so we consider two typical situations: (1) H diffuses around a Si-C tetrahedron, (2) H diffuses between Si-C tetrahedrons. In the pure SiC, H prefers to bond with C at ABc sites or BC sites. So we first concern the diffusion behaviors of H from a site ABc to a site BC around C in a Si-C tetrahedron. The calculated energy profile for H diffusion is shown in Fig.4 (a). From Fig.4(a), we can see that the diffusion barrier is about 0.07 eV for H migrating from the site ABc″ to the site BC, and for the equivalent paths of ABc′-BC′ 、ABc‴-BC′. Reversely, the diffusion barrier is only 0.01 eV for H from site BC to site ABc (i.e. paths of BC-ABc″、BC′-ABc′ and BC′-ABc‴). Such small energy barrier means that H moves easily around the C atom in a Si-C tetrahedron via ABc sites BC sites. Then we consider that an H atom migrates between the Si-C tetrahedrons. Two typical diffusion paths are concerned: one is near the bonds of C-Si, the other is via the interstitial site of Tsi. The related diffusion energy profiles are also plotted in Fig.4 (a). As can be seen, the diffusion barriers are about 0.48 eV and 0.52 eV for the path of ABc-ABc′ and the path of ABc-Tsi, respectively. Meanwhile, the diffusion barriers are about 0.51 eV and 0.46 eV for the path of BC-BC′ and the path of BC(BC′)-Tsi, respectively. Reversely, the diffusion barriers are less than 0.21eV for the path or Tsi-ABc or Tsi-BC. So H can migrate from a Si-C tetrahedron to another by overcoming the barriers of less than 0.52 eV. We can conclude that an H atom can diffuse in the pure SiC materials via the paths of ABc-ABc′-BC′-ABc‴ and ABc-Tsi-BC-ABc″with the energy cost less than 0.48eV and 0.52eV respectively.

We then focus on the diffusion behaviors of H in SiC with the implanted He. In SiC+He, the most stable site for H is the nearest BC sites around He, so we study some typical diffusion pathways of H around He, as shown in Fig.4(b), in which He is at the site Tsi′. In Fig.4(b), sites of BC、BC′、BC‴ and ABc are the nearest typical sites around He, whereas sites of Tsi, BC″ and ABc′ are a little far from He. Different from that in the pure SiC, the diffusion barrier for H from the bond center site (BC, the most stable site for H) to its neighboring site BC (path of BC-BC′) goes up to be 1.0 eV, due to presence of He. Similarly, the energy barrier for H diffusing from site BC′ to its neighboring ABc (path of BC′-ABc) is about 0.43 eV, while that from site BC‴ to site ABc (path of BC‴ -ABc) is as high as 0.98 eV. Meanwhile, the barrier for diffusion path of BC‴-Tsi (Tsi is the neighboring tetrahedral interstitial of four Si from He) is also up to be 0.95 eV. We note that these paths are around He, in which the barriers for H diffusion are much higher than that in the pure SiC. It drops a hint that the implanted He blocks the diffusion of H nearby. Comparing with that at the site BC near He, the formation energy for H at these BC sites a little far from He(i.e. BC") goes up significantly to be about 0.6eV, higher than that at ABc′ site by about 0.1 eV, being similar to that in the pure SiC. This implies the effect He on site preference of H in SiC within a short range, in consistent with the experimental observation that the retained H distribution is nearly not affected by preceding He implantation in 3C-β SiC[14]. Moreover, the diffusion barriers for paths of BC"-Tsi and ABc′-BC" are about 0.46 eV and 0.06eV, respectively, which is almost the same as that of H diffusing in the pure SiC.

Finally we try to understand such significant increment of H diffusion barrier induced by the presented He nearby. Essentially, the height of H diffusion barrier is directly determined by the energy difference between the equilibrium and transition states [24], thus we just need to explore the energies of these states in pure SiC and that in SiC+He. We then take the typical path of BC-BC' in the both pure SiC and SiC+He as an example to address it. For convenience, the equilibrium states are denoted as BC state and BC' state respectively, and the transition state is denoted as TR state. In the previous section, we point out that the reduction of hybridization of C-Si bond caused by the implanted He atom induces the reduction of formation energy of H at BC site. The formation energy of H at BC(BC') state in SiC+He are lower by about 0.48 eV than that in the pure SiC. Differently, we find out that the formation energies of H at TR state is just changed slightly in the SiC+He, as compared to that in the pure SiC. As a result, the energy barrier of BC-BC' path in SiC+He is 0.48 eV higher than in pure SiC. So we speculate that the implanted He mainly influence the Si-C bond in its vicinity to some extent.

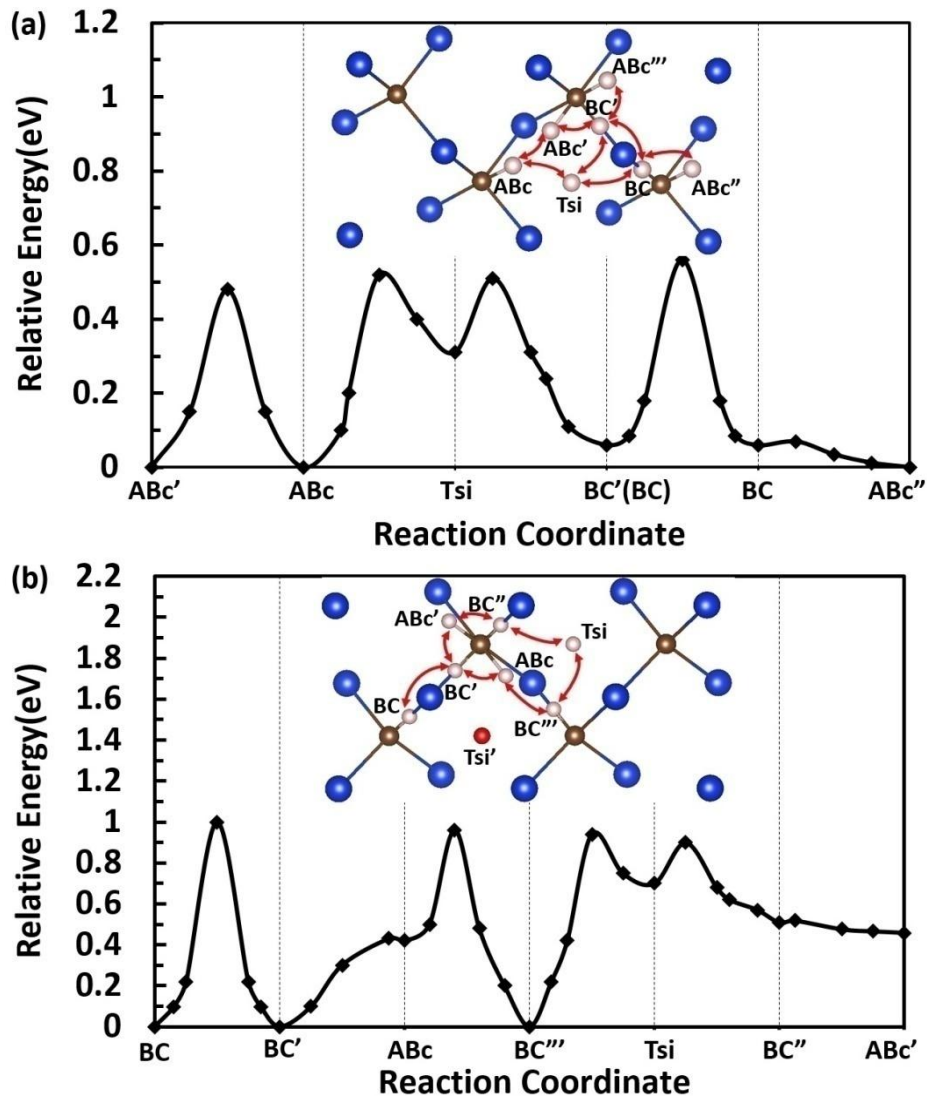

Fig.4 Energy profiles for H diffusing in (a) SiC and (b) SiC+He. Blue, brown, red, pink balls represent Si, C, He, H atoms, respectively.

## 4. Conclusions

By investigating site preference and diffusion behaviors of H in the pure SiC and the He-implanted SiC on the basis of the first-principles calculations, we find that the most stable site for H, the anti-bond site of a Si-C bond at the side of C, in the pure SiC change to be the nearest bond-center site of the Si-C bonds around He in the SiC+He. The analysis on the density of states reveals that such change can be attributed to the weaker hybridization of the C-Si bond caused by the implanted He. Moreover, significant blue shift and red shift for H at ABc site and BC site appear in vibrational spectra of SiC due to the presence of He. In addition, the diffusion barriers of H go up significantly around the implanted He, indicating He blocks the diffusion of H nearby. However, the implanted He in SiC just affect the diffusion of H in a short range, within the nearest neighbor interstitial sites, being agreement with the experimental reports.

## References

[1] L.L. Snead, T. Nozawa, M. Ferraris, Y. Katoh, R. Shinavski, M. Sawan. J. Nucl.Mater. 417 (2011) 330–339.

[2] L. Giancarli, H. Golfier, S. Nishio, A.R. Raffray, C.P.C. Wong, R.Yamada. Fusion Engineering and Design. 61–62 (2002) 307–318.

[3] Y. Katoh, L.L. Snead, I. Szlufarska, W.J. Weber, Curr. Opin. Solid State Mater.Sci. 16 (2012) 143–152.

[4]Yutai Katoh, Kazumi Ozawa, Chunghao Shih, Takashi Nozawa, Robert J. Shinavski, Akira Hasegawa, Lance L. Snead. J. Nucl.Mater. 448 (2014) 448–476.

[5] Q.G. Guo, Z.J. Liu, J.G. Li, N. Noda, Y. Kubota, L. Liu, J. Nucl. Mater. 363 (2007) 1216-1220.

[6] G.N. Luo, Q. Li, M. Liu, X.B. Zheng, J.L. Chen, Q.G. Guo, X. Liu, J. Nucl. Mater. 417 (2011) 1257-1261.

[7] Yan-Ru Lin, Chun-Yu Ho, Wei-Tsung Chuang, Ching-Shun Ku, Ji-Jung Kai. J. Nucl.Mater. 455 (2014) 292–296.

[8] Sosuke Kondo, Takaaki Koyanagi, Tatsuya Hinoki. J. Nucl.Mater. 448 (2014) 487–496.

[9] Yan-Ru Lin, Ching-Shun Ku, Chun-Yu Ho, Wei-Tsung Chuang , Sosuke Kondo,

Ji-Jung Kai. J. Nucl.Mater. 459 (2015) 276–283.

[10] Lance L. Snead, Yutai Katoh, Takaaki Koyanagi, Kurt Terrani, Eliot D. Specht. J. Nucl.Mater. 471 (2016) 92–96.

[11] P. Jung, H. Klein, J. Chen. J. Nucl.Mater. 283-287 (2000) 806-810.

[12] J.H. O'Connell, J.H. Neethling. Radiat. Eff. Defects Solids 167 (2012) 299-306.

[13] Yasuhisa Oya, Hideo Miyauchi, Taichi Suda, Yusuke Nishikawa. Fusion Engineering and Design 82 (2007) 2582–2587.

[14] Alec Deslandes, Mathew C. Guenette, Lars Thomsen, Mihail Ionescu. J. Nucl.Mater. 469 (2016) 187-193.

[15] Yunosuke Ishikawa, Shinji Nagata, Ming Zhao, Tatsuo Shikama. J. Nucl.Mater. 455 (2014) 512–515.

[16] Jingjing Sun, Yu-Wei You, Jie Hou, Xiangyan Li, B.S. Li, C.S. Liu and Z.G. Wang. Nucl. Fusion 57 (2017) 066031 (11pp).

[17] Adrien Couet, Jean-Paul Crocombette, Alain Chartier. J. Nucl.Mater. 404 (2010) 50–54.

[18] A. Gali, B. Aradi and N.T. Son. Phys. Rev. Lett., 81(2000) 4926.

[19] A. Gali and N.T. Son. Phys. Rev. B 71(2005) 035213.

[20] B. Aradi, A. Gali, P. Deak, J. E. Lowther, N. T. Son, E. Janzen, W. J. Choyke. Phys. Rev. B (2001) 63 245202.

[21] A.Charaf Eddin, L. Pizzagalli. J. Nucl.Mater. 429 (2012) 329–334.

[22] M. Van Ginhoven, Alain Chartier, Constantin Meis, William J. Weber. J. Nucl.Mater. 348 (2006) 51–59.

[23] Jong Hyun Kim, Yong Duk Kwon Parlindungan Yonathan, Ikhwan Hidayat. J Mater Sci (2009) 44:1828–1833.

[24] A Yungang Zhou, Qi Liu. Journal of Alloys and Compounds 647 (2015) 167-171.

[25] G. Kresse, J. Furthmüller. Phys. Rev. B 54 (1996) 11169.

[26] P.E. Blöchl. Phys. Rev. B 50 (1994) 17953.

[27] J.P. Perdew, K. Burke, M. Ernzerhof. Phys. Rev. Lett. 77 (1996) 3865.

[28] H.J. Monkhorst, J.D. Pack. Phys. Rev. B 13 (1976) 5188.

[29]Tsutoma Kawamura. Mineralogical Journal 4 (1965) 333-355.

[30] G.H. Lu, H.B. Zhou, C.S. Becquart, Nucl. Fusion 54 (2014) 086001.

[31] G. Mills, H. Jónsson, G.K. Schenter, Surf. Sci. 324 (1995) 305.

[32] Henkelman, G.; Uberuaga, B. P.; Jonsson, H. J. Chem. Phys. 113(2000) 9901.

[33] Henkelman, G.; Jonsson, H. J. Chem. Phys. 113(2000) 9978.

[34]Lei Zhang, Ying Zhang, Guang-Hong Lu. Nuclear Instruments and Methods in Physics Research B 267 (2009) 3087–3089.

[35]M. Kaukonen, C. J. Fall, and J. Lento. Appl. Phys. Lett., 83(2003) 923.

[36] Hass, R. Feng, J. E. Millán-Otoya, X. Li. Phys. Rev. B 75(2007) 214109.

[37] F. Fascale, C. M Zicovich-Wilson, F. Lopez Gejo, B. Civalleri, R. Orlando, R.Dovesi. J. Comput. Chem. 25(2004) 888.